\documentstyle[12pt]{article}
\begin{document}
\def\r{\rightarrow}
\def\be{\begin{equation}}
\def\ee{\end{equation}}
\def\ben{\begin{eqnarray}}
\def\een{\end{eqnarray}}
\baselineskip 21pt plus .1pt minus .1pt
\pagestyle{empty}
\noindent
\begin{flushright}
SINP-TNP-07 \\
March 1999\\
\end{flushright}
\vskip .1in
\begin{center}
{\large\bf{Light Dirac 
Neutrinos 
In An 
$SU(2)_L\times U(1)_Y$ Model}} 
\end{center}
\vskip .25in
\begin{center}
Ambar Ghosal
\end{center}
\vskip .1in
\begin{center}
Saha Institute of Nuclear Physics\\
1/AF, Bidhannagar, 
Calcutta 700 064, India\\
\end{center}
\noindent
Light Dirac neutrino of mass  of mass of the 
order of few eV is obtained in an $SU(2)_L\times U(1)_Y$ 
model with an extended Higgs sector and right-handed neutrinos. 
Small neutrino mass is generated at the tree level 
through small effective coupling of the Dirac neutrino 
mass term due to soft discrete symmetry breaking. 
In order to remove the exact degeneracy in mass between 
the second and the third generation of neutrinos, 
one loop corrected mass terms are incorporated. 
The model can accommodate bi-maximal mixing scenario of 
neutrino which has been favoured by recent Solar and 
atmospheric neutrino experiments.     
\noindent
E-mail: ambar@tnp.saha.ernet.in
\newpage
\pagestyle{plain}
\setcounter{page}{2}
\noindent
The long standing conjecture of 
neutrino oscillation as well as non-zero 
neutrino mass has recently been favoured 
by the result of Atmospheric neutrino 
experiment observed by the Super-Kamiokande 
Collaboration [1].
The results of SOUDAN [2] and CHOOZ [3] 
experiments are also consistent with the 
Super-Kamiokande experimental results. 
The reported result of Super-Kamiokande 
experiment leads to the oscillation 
of $\nu_\mu\r \nu_\tau$ (or $\nu_s$) with 
the mass-squared diference 
$\Delta m^2_{\mu\tau}$ $\sim$ $5 
\times {10}^{-4} - 
6\times {10}^{-3}$ $\rm{eV^2}$ as well as the 
mixing between the neutrino species is maximal. 
Furthermore, the solar neutrino problem could 
be resolved by the $\nu_e\r\nu_\mu$ oscillation 
with 
$\Delta{m^2}_{e\mu}$ $\sim$ $(0.3 - 0.7)
\times {10}^{-5}$ $\rm{eV}^2$ and 
$Sin^22\theta_{e\mu}\sim 
3.5\times {10}^{-3}$ if the small angle MSW 
solution is considered. The large angle MSW 
solution predicts the value of 
$\Delta {m^2}_{e\mu}\sim {10}^{-5}-{10}^{-4}$ 
$\rm{eV^2}$ and 
$Sin^22\theta_{e\mu}\sim$ 0.8 - 1 [4].     
\par
In order to reconcile simultaneously 
the solar and the 
atmosheric neutrino problem as well as 
the candidature of neutrino as a hot 
dark matter component in a mixed dark 
matter scenario it has been pointed out 
[5] that the three electroweak neutrinos 
are almost degenerate in mass of the 
order of few eV. Explicit realization 
of almost degenerate neutrino mass has been 
demonstrated by several authors [5,6].  
Furthermore, the present solar and 
atmospheric neutrino experimental 
data suggests the bi-maximal mixing pattern
[7] that is maximal mixing between $\nu_e$ 
-$\nu_\mu$ and $\nu_\mu$ - 
$\nu_\tau$.
\par
In the present work, we demonstrate that an 
$SU(2)_L\times U(1)_Y$ model with discrete 
$S_3\times Z_3\times Z_4$ symmetry, 
right-handed neutrinos and appropriate 
Higgs fields can give rise to almost 
degenerate Dirac neutrino of  mass   
of the order of few eV 
as well as 'bi-maximal' mixing 
pattern in the neutrino- charged lepton 
charged current interactions. The scenario of 
degenerate Dirac neutrino  has recently been 
investigated [8] in view of Super-Kamiokande 
experimental result. In general, the 
Dirac neutrinos are much heavier 
than the Majorana 
neutrinos unless by some mechanism the Yukawa 
couplings associated with the Dirac 
neutrino mass terms
are made smaller. In the present work, Dirac 
neutrino of mass of the order of few 
eV is generated 
through the small effective coupling 
of the Dirac 
neutrino mass term. The methodology has 
been proposed 
in Ref.[9] in order to generate small 
Majorana neutrino 
mass in the context of an 
$SU(2)_L\times U(1)_Y$ model. 
The present model gives rise 
to light Dirac neutrino 
mass at the tree level as well 
as at the one loop level 
which can accommodate the solar 
and atmospheric neutrino 
experimental results. We have 
discarded the LSND observed 
neutrino experimental result [10] 
due to its mismatch 
with the KARMEN [11] neutrino experimental result.    
\par
\noindent
We concentrate on the lepton 
and Higgs fields of the model. 
The lepton
content in the present model 
is as follows\hfill
\ben
l_{iL}\;\; (2,\; -1,\; 1), & 
\nu_{iR}\;\; (1,\; 0,\; 1)
\een
where $i$ = 1, 2, 3 is the 
generation index. 
The first two digits in the 
parenthesis 
respectively
represent $SU(2)_L$ 
and $U(1)_Y$ quantum numbers 
and 
the last 
digit 
represents lepton 
number 
${\rm{L}}(= {\rm{L}}_e + 
{\rm{L}}_\mu 
+ {\rm{L}}_\tau)$ 
which is conserved in the present 
model.
The following Higgs fields
are considered with the 
vacuum expectation values 
(VEV's) as indicated\hfill
\ben
\phi_i \;(2, \; 1,\; 0)\;\; =
\;\; \pmatrix{0\cr v_i}, 
& \eta\;\;
(1,\; 0,\; 0)\;\; = \;\; k
\een  
where $i$ = 1,$\ldots$ 4 
are the 
number of doublet 
Higgs fields, and $\eta$ is a real 
scalar singlet.  
The 
singlet Higgs 
fields do not couple with 
the leptons
at the tree level due to 
the discrete symmetry 
incorporated in the model.\hfill
\par
The lepton and Higgs 
fields transform under 
discrete $S_3$ $\times$ 
$Z_3\times$ $Z_4$ symmetry 
as follows:\hfill
\vskip .25in
\noindent
i)$\;\;$ {\underline{$S_3$ 
\rm{Symmetry}}}\hfill 
\ben
(l_{2L}, l_{3L})\r 2,\;\; 
l_{1L}\r 1,\;\; 
(\nu_{\tau R}, 
\nu_{\mu R})\r 2\nonumber\\
\nu_{e_R}\r 1, 
(e_R \, , \, \mu_R) \r 2, 
\tau_R\r 1,
\nonumber\\ 
\phi_1\r 1, \phi_2\r 1,
(\phi_3, \phi_4)\r 2, 
\eta\r 1 
\een
\vskip .25 in
\noindent
ii)$\;\;$ {\underline{
$Z_3\times$$Z_4$ 
\rm{Symmetry}}}\hfill 
\ben
(l_{2L}, l_{3L})\r 
\omega^\star(l_{2L}, l_{3L}), 
l_{1L}\r l_{1L}, 
(\nu_{\tau R}, \nu_{\mu R})\r 
\omega^\star(\nu_{\tau R},\nu_{\mu R}),
\nonumber\\
\nu_{e R}\r \nu_{e R}, 
(e_R\, , \, \mu_R)\r 
-i\omega^\star(e_R\, , \, \mu_R), 
\tau_R\r -i\omega^\star\tau_R,\nonumber\\
\phi_1\r\phi_1,\phi_2\r i\phi_2,
(\phi_3, \phi_4)\r 
i\omega(\phi_3, \phi_4),\nonumber\\
\eta\r -\eta 
\een
where $\omega$ = $exp(2\pi i/3)$
\vskip .25in
\noindent
The discrete symmetry  
 gives 
rise to some vanishing
elements in the Dirac 
neutrino 
mass matrix and the charged lepton mass 
matrix at the 
tree level. 
The purpose of 
incorporation of 
$S_3$ 
permutation
symmetry  
is to generate the 
equality 
between the Yukawa 
couplings
associated with the 
neutrinos in 
order to get degenerate 
neutrino 
mass. The exact degeneracy is lifted 
due to the incorporation of one loop 
corrected neutrino mass.
The Higgs field $\phi_1$ 
couples 
only with the neutrinos 
and is prohibited 
from coupling with the 
charged leptons. 
We will estimate the value 
of 
$v_1$ by utilizing the 
minimization
condition of the Higgs 
potential. 
The $\phi_2$ Higgs field 
couples only with the 
charged leptons as well as
gives rise to small 
neutrino 
mass through its coupling 
with 
$\phi_1$ and  $\eta$ 
Higgs fields through soft discrete 
symmetry breaking term contained in 
the Higgs potential of the present model. 
The purpose of 
incorporation
of $\phi_3$ and $\phi_4$ 
Higgs fields is to achieve 
non-degenerate charged 
lepton 
mass matrix. 
Apart from the 
electroweak symmetry 
breaking scale, the 
present 
model
contains another 
intermediate symmetry 
breaking scale at 
which the 
non-zero
VEV of $\eta$    
Higgs fields is developed. The smallness of 
the 
neutrino mass is 
controlled by the
VEV of the singlet 
Higgs field and the 
coefficient of the 
soft discrete symmetry violating term 
relating $\phi_1$, 
$\phi_2$, $\eta$  
Higgs fields 
contained in the Higgs 
potential. The soft discrete symmetry 
breaking term is necessary to remove any 
extra U(1) global symmetry in the model. 
We also discard any hard discrete symmetry 
breaking term for our analysis.
\hfill
\vskip .25in
\par
The most general renormalizable, 
Higgs potential in the present 
model can be
expressed as \hfill
\be
V = V(\phi_i) + V(\eta) + 
V(\phi_i, \eta_j).
\ee
where $V(\eta)$ is not 
relevant for the present 
analysis and
the rest of the terms are 
explicitly given by 
\be
V(\phi_i) =  
m_1^2(\phi_1^\dagger\phi_1) 
-m_2^2(\phi_2^\dagger\phi_2) +
\sum_{i=1}^{2} 
\mu_i
{(\phi_i^\dagger\phi_i)}^2 
-m_3^2(\phi_3^\dagger\phi_3 
+\phi_4^\dagger\phi_4) +$$ 
$$\mu_3
{(\phi_3^\dagger\phi_3 +
\phi_4^\dagger\phi_4)}^2 +
\lambda_1
(\phi_1^\dagger\phi_1)
(\phi_2^\dagger\phi_2)$$
$$+\lambda_2
(\phi_1^\dagger\phi_1)
(\phi_3^\dagger\phi_3 + 
\phi_4^\dagger\phi_4)
+\lambda_3
(\phi_2^\dagger\phi_2)
(\phi_3^\dagger\phi_3 
+\phi_4^\dagger\phi_4)$$                       
$$+ 2\lambda_4
(\phi_3^\dagger\phi_4
\phi_4^\dagger\phi_3)             
\ee
and
\be
V(\phi_i, \eta) = 
\sum_{i=1,..4}
\lambda_{ii}
(\phi_i^\dagger\phi_i)
(\eta^2)\, 
({\rm{with}}\,  
\lambda_{33}=\lambda_{44})
+\lambda^\prime(\phi_1^\dagger
\phi_2\eta + 
\phi_2^\dagger\phi_1\eta)               
\ee
\noindent
It is to be noted that, in 
order to get neutrino mass 
$\sim$ eV, we consider 
the VEV of $\phi_1$ is zero 
at the tree level. 
This has been 
achieved by choosing 
positive mass 
term $(m_1^2>0)$ of the 
$\phi_1$
Higgs field in Eq.(6). 
The non-zero 
VEV of $\phi_1$, arising 
due to the presence   
of the soft discrete symmetry breaking 
term $\lambda^\prime$, is 
estimated as follows.
Substituting the VEV's of 
the Higgs 
fields in Eqs.(6) and (7) 
and 
minimizing the entire Higgs 
potential with respect to 
$v_1$, we get
\begin{equation}
v_1 = -{B\over A}
\end{equation}
\noindent
with 
\be
B= \lambda^\prime v_2 k 
\ee 
\be
A = m_1^2 + \lambda_1 v_2^2 
+ \lambda_2(v_3^2 + v_4^2)
+ \lambda_{11}k^2  
\ee
\noindent
where we have neglected 
$\mu_1$ 
term for simplicity. 
On simplification of 
Eq.(8), 
we obtain
\begin{equation}
v_1 =-{{\lambda^\prime 
v_2 k }
\over m_1^2}
\end{equation}
\noindent
assuming $m_1^2$ to be 
much larger than all other 
terms 
in Eq.(10). However, the 
above
assumptions do not affect 
the essential results 
derived in
our present analysis.\hfill
\vskip .25in
The most general 
discrete symmetry invariant 
lepton-Higgs Yukawa 
interaction, 
in our present model is as 
follows\hfill
\ben
L_Y & = &[f_1
(\bar{l_{2L}}\nu_{\tau R} 
+ \bar{l_{3L}}\nu_{\mu R})
+ f_2 \bar{l_{1L}}\nu_{e R}]
\tilde{\phi_1} + 
\nonumber\\
& + &
g_1(\bar{l_{2L}} e_R + 
\bar{l_{3L}}\mu_R)\phi_2+
g_2(\bar{l_{2L}} \phi_3 + 
\bar{l_{3L}}\phi_4)\tau_R
\nonumber\\
& + & g_3\bar{l_{1L}}(e_R\phi_3 + \mu_R\phi_4) 
+H.c.
\een
Substituting the VEV's of 
the Higgs fields in Eq.(12) 
the tree level Dirac neutrino mass 
matrix $M_{\nu}^0$ comes out as \hfill
\be
M_\nu^0 = \pmatrix{\xi a&0&0\cr
                   0&0&a\cr
                   0&a&0}
\ee
where 
\be
a=f_1 v_1, \xi = {f_2\over f_1}
\ee
\noindent
In order to 
remove the exact degeneracy between the mass of 
the second generation and the third generation , 
we have incorporated one-loop mass terms arising 
due to the charged Higgs exchange. The one-loop corrected 
neutrino mass matrix is given by \hfill
\be
M_\nu^\prime = \pmatrix{x_1 & 0 & x_2\cr
                        x_3 & x_4 & 0\cr
			0 & x_5 & x_6}
\ee
where \hfill
\be
x_1 = g_3 f_2 \lambda_2 v_3 v_1 m_e F(M_3^2, M_1^2)
\ee
\be
x_2= g_3 f_1 \lambda_2 v_4 v_1 m_\mu F(M_4^2, M_1^2)
\ee
\be
x_3 = g_1 f_2(\lambda^\prime k 
+ \lambda_1 v_1 v_2)m_e F(M_2^2, M_1^2)
\ee
\be
x_4 = g_2 f_1 \lambda_2 v_3 v_1 m_\tau F(M_3^2, M_1^2)
\ee
\be
x_5 = g_2 f_1 \lambda_2 v_1 v_4 m_\tau F(M_4^2, M_1^2)
\ee
\be
x_6 = g_1 f_1 (\lambda^\prime k + 
\lambda_1 v_1 v_2)m_\mu F(M_2^2, M_1^1)
\ee
and \hfill
\be 
F(M_i^2, M_j^2) = {\frac{1}{16\pi^2 (M_i^2 - M_j^2)}}
\rm{ln}\frac{M_i^2}{M_j^2}   
\ee
and $M_i's $ are the masses of the charged Higgs fields. 
Combining Eq.(13) and (15), we get the total mass matrix of 
the neutrino as \hfill
\be
M_\nu = M_\nu^0 + M_\nu^\prime  
\ee
The charged lepton mass matrix comes out from Eq.(12) 
as\hfill
\be
M_l = \pmatrix{g_4 v_3& g_4 v_4 & 0 \cr
               g_2 v_2 & 0  & g_3 v_3\cr
               0 & g_2 v_2 & g_3 v_4}
\ee
In this situtation, it is not possible to diagonalise  		  
simultaneously the neutrino mass matrix and the charged 
lepton mass matrix and the mismatch between the 
diagonalisation 
matrices will give rise to the CKM -type mixing 
matrix in the leptonic sector.
Before 
going to diagonalise the neutrino mass matrix $M_\nu$ 
, we assume the following to simplify the diagonalisation 
procedure without altering any essential results. First 
of all, we neglect all the terms in $M_\nu^\prime$ 
proportional to $m_e$ and $m_\mu$ since these are 
small compared to the terms proportional to $m_\tau$. 
We consider the weak eigenstates and the mass eigenstates 
are related by the following relation\hfill
\be
\pmatrix{e\cr
	\mu\cr
	\tau}
= U_l \pmatrix{l_1\cr
	       l_2\cr
	       l_3}
\ee
where
\be
U_l =\pmatrix{c_{12}c_{13}&s_{12}c_{13}&s_{13}\cr
              -s_{12}c_{23}-c_{12}s_{23}s_{13} &
               c_{12}c_{23}-s_{12}s_{23}s_{13}&
               s_{23}c_{13}\cr
               s_{12}s_{23}-c_{12}c_{23}s_{13}& 
               -c_{12}s_{23}-s_{12}c_{23}s_{13}&
               c_{23}c_{13}}
\ee
and $c_{ij} = Cos\theta_{ij}$, 
$s_{ij} = Sin\theta_{ij}$ , i,j =1,2,3 
are the generation indices. A similar relation for 
the neutrino sector is also considered  	
\be
\pmatrix{\nu_e\cr
	 \nu_\mu\cr
	 \nu_\tau}
=U_\nu \pmatrix{\nu_1\cr
                \nu_2\cr
                \nu_3}
\ee
where 
\be
U_\nu = \pmatrix{c_{12}^\prime c_{13}^\prime
                 & s_{12}^\prime c_{13}^\prime &
                 s_{13}^\prime\cr
                 -s_{12}^\prime c_{23}^\prime 
                -c_{12}^\prime s_{23}^\prime s_{13}^\prime &
                 c_{12}^\prime c_{23}^\prime 
                - s_{12}^\prime s_{23}^\prime s_{13}^\prime &
                s_{23}^\prime c_{13}^\prime\cr
                s_{12}^\prime s_{23}^\prime 
                -c_{12}^\prime c_{23}^\prime s_{13}^\prime &
                -c_{12}^\prime s_{23}^\prime 
               -s_{12}^\prime c_{23}^\prime s_{13}^\prime & 
               c_{23}^\prime c_{13}^\prime}
\ee
and $c_{ij}^\prime$ = $Cos\theta_{ij}^\prime$ 
, $s_{ij}^\prime$ = $Sin\theta_{ij}^\prime$. In order to obtain 
'bi-maximal' scenario , we set the mixing angles as\hfill
\be
\theta_{13} = 0 , \theta_{23} = {\frac{\pi}{8}} , 
\theta_{12} = {\frac{\pi}{4}}   
\ee
for the charged lepton sector and 
\be
\theta_{13}^\prime = 0 , \theta_{23}^\prime =\frac{\pi}{8}, 
\theta_{12}^\prime = 0
\ee
for the neutrino sector. The zero value's of 
$\theta_{13}$,  $\theta_{13}^\prime$ 
and $\theta_{12}^\prime$ 
is obvious from the structure of the mass matrices, 
and we set the non-zero values by adjusting model 
parameters.  
The above choice of   
mixing angles give rise to the following mixing 
matrix V in the neutrino-charged lepton charged current 
interaction which is given by 
\be 
V= U_\nu^\dagger U_l 
 = \pmatrix{\frac{1}{\sqrt 2}& 
   \frac{1}{\sqrt 2}& 0\cr
           -\frac{1}{2}& \frac{1}{2} 
            & \frac{1}{\sqrt 2}\cr 
            \frac{1}{2} & -\frac{1}{2} & \frac{1}{\sqrt 2}}
\ee
as desired to obtain ' bi-maximal' mixing scenario to explain 
solar and atmospheric neutrino experimental data. 
On diagonalisation 
of $M_\nu M_\nu^\dagger$, we obtain the following 
eigenvalues\hfill
\be 
m_{\nu_1}^2 = \xi^2 a^2
\ee
\be
m_{\nu_2}^2 = \frac{(q+s) 
+ \sqrt{{(q-s)}^2+ 4 r^2}}{2}
\ee
\be
m_{\nu_3}^2 = \frac{(q+s) 
- \sqrt{{(q-s)}^2 + 4r^2}}{2}
\ee
\noindent
where $q = a^2 + x_4^2$, $s = {(a + x_5)}^2$, 
$r = x_4{(a + x_5)}$ and the mixing angle 
$\theta_{23}^\prime$ is given by\hfill
\be
tan2\theta_{23}^\prime = \frac{2r}{s-q}
\ee
\noindent
Neglecting higher orders of $x_4$ and $x_5$, 
and assuming $M_4\sim M_3$ and $M_4,M_3$ $>>M_1$, 
the mass-squared differences 
$m_{\nu_2}^2$ $- m_{\nu_3}^2$, 
$m_{\nu_2}^2$ $- m_{\nu_1}^2$ and the mixing angle 
$\theta_{23}^\prime$, can be simplified as\hfill      
\be
m_{\nu_2}^2 - m_{\nu_3}^2 \sim 2 a x_4
\ee
\be
m_{\nu_2}^2 - m_{\nu_1}^2 \sim a^2(1 - \xi^2)
\ee
\be
tan2\theta_{23}^\prime \sim \frac{v_3}{v_4}
\ee
In order to accommodate the atmospheric neutrino 
experimental data, we set the mass-squared difference 
at a typical value as      
$m_{\nu_2}^2 - m_{\nu_3}^2\sim 5\times {10}^{-4}$
 and $tan 2\theta_{23}^\prime\sim 1$. For a numerical estimation 
, we consider the following choices of model parameters
as $M_4\sim M_3 = 400 \rm{GeV}$, 
$M_1\sim M_2 = 200 \rm{GeV}$, $v_2\sim v_3\sim v_4  
= 100$ GeV, $k = 1$ TeV and $m_1 = 1$ TeV. 
The above choice of model parameters gives rise to 
the mass-squared 
difference as required to solve the atmospheric neutrino 
problem with the constraint on the couplings as 
$f_1^2\lambda^{{\prime}^2} 
g_2\lambda_2\sim 0.5\times {10}^{-3}$. 
The parameter $\xi$ is estimated from the solar neutrino 
experimental data and for 
$m_{\nu_2}^2 - m_{\nu_1}^2$ $\sim$ ${10}^{-5}$ 
(for large angle MSW solution), the parameter $\xi$ 
comes out slightly less than unity.
\par
In summary, we have demonstrated that an 
$SU(2)_L\times U(1)_Y$ model with 
$S_3\times$ $Z_3\times$ $Z_4$ discrete symmetry, right-handed 
neutrinos and appropriate Higgs fields give rise to 
light Dirac neutrino of mass of the order of    
few eV. Small neutrino mass is generated at the tree 
level due to the small effective coupling of the 
Dirac neutrino mass terms 
through the incorporation of soft discrete symmetry 
breaking term. In order to remove the 
exact degeneracy in mass between the second and the 
third generation of neutrinos, we have incorporated 
one loop corrected mass terms. The present model 
accommodates the  
solar and atmospheric neutrino experimental 
result  through bi-maximal 
mixing angle  
scenario 
with a reasonable choice
of model parameters.
\vskip .25in
\noindent 
Author  acknowledges Utpal Sarkar, Debajyoti Choudhury, Anirban Kundu, 
Biswarup 
Mukhopadhyaya and Sourov Roy 
for many helpful comments and discussions. 
\newpage
\begin{center}
{\large\bf{References}}
\end{center}
\begin{enumerate}
\item T. Kajita, Talk in 'Neutrino 98', Takayama, 1998, 
Super-Kamiokande Collaboration,  
Y. Fukuda et al., hep-ex/9805006, 
hep-ex/9805021, hep-ph/9807003. 
\item S.M.Kasahara et al., Phys. Rev. D55 (1997) 5282.
\item M. Appollonio et al. Phys. Lett. B420 (1998) 397.
\item J. Bahcall, P. Krastev and A. Yu. Smirnov, 
hep-ph/9807216.
\item D. O.Caldwell and R. N. Mohapatra, 
Phys. Rev. D50, (1994) 3477, A.S.Joshipura, 
Z.Phys. C 64 (1994) 31. 
\item A.S.Joshipura, Phys. Rev. D51(1995), 1321, 
D. G. Lee and R. N. Mohapatra, Phys. Lett. B229, 
(1994) 463, P.Bamert and C.P.Burgess, Phys. Lett. B329, 
(1994) 289; A. Ioannissyan and J.W.F. Valle, ibid, 
332, (1994) 93, A. Ghosal, Phys.Lett. B398, (1997) 315, 
A. K. Ray and S. Sarkar, Phys. Rev. D58, (1998) 055010.
\item V. Barger, S. Pakvasa, T. J. Weiler and  K. Whisnant, 
hep-ph/9806387, R. N. Mohapatra and S. Nussinov, 
hep-ph/9808301, K. Kang, S. K. Kang, C. S. Kim and  
S. M. Kim, 
hep-ph/9808419, S. Mohanty, D. P. Roy and  U. Sarkar,   
hep-ph/9808451, B. Brahmachari, hep-ph/9808331. 
\item U.Sarkar, hep-ph/9808277. 
\item G.Gelmini and T.Yanagida, Phys. Lett. B294, (1992) 53.
\item LSND Collaboration: C.Athanassopoulos et al., 
Phys. Rev. Lett. 75 (1995) 2650, Phys. Rev. C54, 
(1996) 2685, Phys. Rev. Lett. 77 ,(1996) 3082, 
nucl-ex/9706006, nucl-ex/9709006, Phys. Rev. Lett.  
81, (1998) 1774.
\item KARMEN Collaboration: R. Armbruster et al, 
Phys. Rev. C57 (1998) 3414, Phys. Lett. B423, (1998) 
15, K. Eitel et al. hep-ex/9809007.
\end{enumerate}
\end{document}